\documentclass[11pt,a4paper]{article}

\usepackage{natbib}
\usepackage{graphicx}
\usepackage{amsmath, amssymb}

\usepackage{tikz}

\newcommand{\A}{\mathbf{A}}
\newcommand{\x}{\mathbf{x}}
\newcommand{\R}{\mathcal{R}}
\newcommand{\high}{\mathrm{high}}
\newcommand{\low}{\mathrm{low}}
\newcommand{\maxi}{\mathrm{max}}
\newcommand{\mini}{\mathrm{min}}
\newcommand{\intervx}[1]{\mathcal{I}_{x_{#1}}}
\newtheorem{definition}{Definition}
\newtheorem{observation}{Observation}
\newtheorem{prop}{Proposition}

\title{Structural requirements and discrimination of cell differentiation networks}

\author{Christian Breindl\footnote{Both authors contributed equally to this work.}, 
  Daniella Schittler\footnotemark[1], Steffen Waldherr, and Frank Allg{\"o}wer\\
  \{breindl,schittler,waldherr,allg{\"ower}@ist.uni-stuttgart.de\} \\[3ex]
  Institute for Systems Theory and Automatic Control,\\ University of
  Stuttgart, 70770 Stuttgart, Germany}

\begin{document}

\maketitle
  

  \begin{abstract}                
    Mathematical models of stem
    cell differentiation are commonly based upon the concept of
    subsequent cell fate decisions, each controlled by a gene
    regulatory network.  These networks exhibit a multistable behavior
    and cause the system to switch between qualitatively distinct
    stable steady states.  However, the network structure of such a
    switching module is often uncertain, and there is lack of
    knowledge about the exact reaction kinetics.  In this paper, we
    therefore perform an elementary study of small networks consisting
    of three interacting transcriptional regulators responsible for
    cell differentiation: We investigate which network structures can
    reproduce a certain multistable behavior, and how robustly this
    behavior is realized by each network.  In order to approach these
    questions, we use a modeling framework which only uses qualitative
    information about the network, yet allows model discrimination as
    well as to evaluate the robustness of the desired multistability
    properties. We reveal structural network properties which are
    necessary and sufficient to realize distinct
      steady state patterns required for cell differentiation.
    Our results
    also show that structural and robustness properties of the
    networks are related to each other.
     \end{abstract}



\section{Introduction} \label{sec:intro}

Stem cells and their potential to give rise to multiple cell types
have become a focus in systems biology and mathematical modeling
throughout the last years \citep{Peltier2010, MacA2009}. Starting
from a multi-potent stem cell state, they undergo the process of cell
differentiation which is completed when the cell ends up in a mature
cell state.  The eventual goal in stem cell research is to guide the
differentiation of stem cells toward a desired cell type safely, and
to maintain this state robustly.  Therefore, a promising yet
challenging task is to reveal the processes driving cell
differentiation by mathematical modeling.

Cell differentiation is commonly viewed and modeled as a sequence of
cell fate decisions \citep{Foster2009}. Thereby, the individual
decision steps can be represented by modules, each driven by a small
gene regulatory network of ``master'' transcriptional regulators (TRs)
in which feedback motifs play a key role
\citep{Xiong2003,Huang2007,Foster2009}.  The development of models for
these modules however is often severely hampered by the lack of
detailed knowledge about molecular processes, shifting the focus
toward qualitative models in order to understand the fundamental
mechanisms.  A variety of publications has contributed to the modeling
of (stem) cell differentiation, covering a broad range of diverse
modeling approaches,
e.g. \cite{Roeder2006,Huang2007,Narula2010,MacA2008}.  Usually,
specific stem cell systems such as hematopoietic or mesenchymal stem
cells are considered. Yet, similar motifs and system properties can be
found among them.  For example, all these approaches have in common
that they represent distinct cell types by distinct stable steady
states, declaring multistability as a universal system property.

One of several aspects within this field concerns the question of
stable cell states, and minimalistic models that are capable of
reproducing experimentally observed cell states.  Although it has been
established to view cell differentiation as subsequent decisions
driven by multistable switching modules \citep{Foster2009}, the
network structure of the single modules is often still unclear.  We
therefore perform an elementary study of small networks consisting of
three TRs. These networks should be able to exhibit one progenitor
state and two competing differentiated cell types. We also compare two
different hypotheses about the steady state levels of the TRs in the
individual cell types.  In order to reveal properties of 3-node
networks that provide candidate models for the differentiation
process, we address the following questions:
\begin{itemize}
\item Given three cell type-specific TRs, which networks of
  interactions between the TRs can reproduce the required stable
  steady states?
\item Among the networks that fulfill the stable steady state
  requirements, are there differences in robustness of
    the required multistability properties with respect to perturbations
    in the interactions?
\end{itemize}
As these are very basic questions, this work builds upon a qualitative
modeling framework which needs only very few assumptions about the
reaction kinetics.  In order to model the influence of a TR on
its target, monotonous activation and inhibition functions are used,
but no exact kinetic parameters have to be specified. This framework
has been previously used for example in \cite{Chaves2008,Breindl2010}.

The outline of this paper is as follows: In Section~\ref{sec:meth}, we
describe the modeling framework, and define several criteria used to
classify the investigated networks. In Section~\ref{sec:res}, we
present and discuss the results of the model analysis.
Section~\ref{sec:con} gives a short summary and concludes with an
outlook on future investigations.

\section{Methods} \label{sec:meth}

The modeling framework used for this work is briefly recalled first.
We then specify the stable steady states to be reproduced by each candidate model, and introduce some characteristic measures and a robustness measure.

\subsection{Modeling framework} \label{sec:meth:mod} 

The framework is based on ordinary differential equations and makes use of only
very few modeling assumptions which shall be explained next.
As we are considering networks of TRs, the influence of
such a regulator on its target gene has to be described mathematically. It is assumed
that this influence can be described by monotonous activation and inhibition functions.
The definitions of these functions are given next.

\begin{definition}
  An \emph{activation (inhibition) function}
  is a function $\nu:[0, \infty) \rightarrow [0,N)$ ($\mu: [0, \infty)
  \rightarrow (0,N]$) with $N \in \mathbb{R}_+$ and:
 \begin{enumerate}
 \item[i)] $\nu$ ($\mu$) is continuously differentiable,
 \item[ii)] $\nu(0) = 0$ and $\nu(x) \rightarrow N$ as $x \rightarrow \infty$ \\
   ($\mu(0) = N$ and $\mu(x) \rightarrow 0$ as $x \rightarrow
   \infty$),
 \item[iii)] $\nu(x)$ ($\mu(x)$) is monotonously increasing
   (decreasing).
 \end{enumerate}
\end{definition}
 
Note, that commonly used kinetics such as Michaelis-Menten or Hill
kinetics are covered by this definition.  Furthermore, it is assumed
that each of the TRs undergoes a linear degradation. With
this, the dynamics of a network of $n$ TRs are given by
\begin{equation}
  \label{eq:system}
  \dot{x}_i = -k_i \cdot x_i + f_i(x), \quad i=1,\ldots,n.
\end{equation}
In this, $x \in \mathbb{R}^n_+$ is the vector of concentrations of 
the TRs and $f_i(x)$ models the combined influence of all TRs
in a network on the TR $x_i$. In this study we only allow multiplicative combinations
of activation and inhibition functions. 
As generally the parameters $k_i$ and the exact shapes of the activation and
inhibition functions are not known, (\ref{eq:system}) only captures
the interaction structure.

As mentioned in the introduction, our goal is to study all possible
networks consisting of three nodes $x_i$, $i \in \{1,2,3\}$. The interaction
structure of such a network is represented by an interaction matrix 
\begin{equation}
  \label{eq:interamat}   
  \A = \left\{a_{ij}\right\}_{i,j \in \{1,2,3 \} },\ a_{ij} := \left\{ 
    \begin{array}{r l} 
      1  & \iff \mbox{$ x_j$ activates $x_i$}\\
      -1 & \iff \mbox{$ x_j$ inhibits $x_i$}\\
      0  & \iff \mbox{no interaction}.
    \end{array} \right.
\end{equation}

To give an example, the interaction matrix
\begin{equation*}
  \A = \left[\begin{array}{ccc}%
      1&0&1\\
      0&1&-1\\
      -1&0&0 \end{array} \right]
\end{equation*}
encodes the network
\begin{equation*}
  \begin{aligned}
    \dot{x}_1 = & -k_1 \cdot x_1 + \nu_1(x_1) \cdot \nu_2(x_3),\\
    \dot{x}_2 = &  -k_2 \cdot x_2 + \nu_3(x_2) \cdot \mu_4(x_3),\\
    \dot{x}_3 = &  -k_3 \cdot x_3 + \mu_5(x_1),
  \end{aligned}
\end{equation*}
in which the parameters $k_i$ and the shapes of the activation and inhibition
functions are not specified.

\subsection{Specifications of stable steady states} \label{sec:meth:speci}

We are specifically interested in the stable steady states of these
networks as they determine the cell type. However, in this context we
will not consider stable steady states as individual points but as
forward-invariant sets in the state space to account for biological
variability. By doing so, only high and low levels of a TR are
distinguished. To be more precise, we assume concentrations
$x_i^\low$, $x_i^\high$ and $x_i^\maxi$ with $0 \leq x_i^\low \leq
x_i^\high \leq x_i^\maxi$ to be known, and a concentration $x_i$ in
the interval $\intervx{i}^\low = [0,x_i^\low]$ is considered as
low. Equivalently, a concentration $x_i \in
\intervx{i}^\high=[x_i^\high,x_i^\maxi]$ is considered as high.  With
this, a stable steady state of (\ref{eq:system}) is considered
as a forward-invariant hyperrectangular set $\x = \intervx{i}^{l_1}
\times \ldots \times \intervx{n}^{l_n}$ with $l_i \in
\{\high,\low\}$. For ease of notation, the n-tuple of labels
$(l_1,\ldots,l_n)$ will be used to refer to this hyperrectangular set.

Let us now turn to our system of three TRs that provide a decision
module in the cell differentiation process. In this setup, three
steady states are of special interest: The progenitor state $A$, a
differentiated cell type $B$, and a competing differentiated cell type
$C$.  Each of these cell types $i \in \{A,B,C\}$ is characterized by a
specific 3-tuple of high and low TR-levels $\x^{(i)} = (l_1^{(i)}, l_2^{(i)},
l_3^{(i)})^T$.  One can however find several plausible ways to characterize
the three different cell types via levels of the three TRs. We assume
the differentiated cell types $\x^{(B)}, \x^{(C)}$ to be characterized
by a high level of the type-specific TR $x_2$ or $x_3$, respectively.
Regarding the progenitor state $\x^{(A)}$, we aim to compare two hypotheses
(``specifications'' of stable steady states, in the following for
short SSS-specifications) against each other, which represent
conceptionally distinct mechanisms:
\begin{itemize}
\item[(S1)] $x_1$ corresponds to a \textit{progenitor factor},
  maintaining the progenitor state. $x_1$ is high in the progenitor
  state $\x^{(A)}$, whereas it is low in the differentiated states
  $\x^{(B)}, \x^{(C)}$. It has to be \textit{down}regulated in order
  to achieve cell differentiation. This means the required stable
  steady states are:
\end{itemize}

\begin{equation}
\begin{aligned}
  \x^{(A)} & := \mathrm{(high, low, low)}, \\
  \x^{(B)} & := \mathrm{(low, high, low)}, \\
  \x^{(C)} & := \mathrm{(low, low, high)}.  \label{eq:specSSs1}   
\end{aligned}
\end{equation}

\begin{itemize}
\item[(S2)] $x_1$ corresponds to a \textit{differentiation factor},
  enabling differentiation. $x_1$ is low in the progenitor state
  $\x^{(A)}$, whereas it is high in the differentiated states
  $\x^{(B)}, \x^{(C)}$. It has to be \textit{up}regulated in order to
  achieve cell differentiation. This corresponds to requiring the
  stable steady states:
\end{itemize}

\begin{eqnarray}
\begin{aligned}
  \x^{(A)} & := \mathrm{(low, low, low)}, \\
  \x^{(B)} & := \mathrm{(high, high, low)}, \\
  \x^{(C)} & := \mathrm{(high, low, high)}.  \label{eq:specSSs2}   
\end{aligned}
\end{eqnarray}

Besides the specified stable steady states, we impose a further
requirement on the candidate models: In order to investigate ``truly
interacting'' nodes rather than assemblages of isolated nodes or
subgraphs, we require each 3-node network to be weakly connected.

\subsection{Characteristic measures} \label{sec:meth:char}

Several characteristic measures are introduced to classify networks
with respect to their interactions, including self-loops, and the
types of these interactions (activating / inhibiting).  The
\textit{number of nonzero entries} in the interaction matrix $\A$,
which equals the zero-norm $|| \A ||_0$, is used as a measure of the
network's connectivity.  To measure the preponderance of activating
versus inhibiting interactions, we use the number of positive entries
in $\A$ minus the number of negative entries in $\A$, which we call
the \textit{positive-negative weight}.

The \textit{number of sign-inconsistent (sign-consistent) loops} is
used to measure the degree of inconsistencies (consistencies,
respectively) arising from network interactions. A loop is either a
feedback loop from $x_i$ to itself, or a feedforward loop from $x_i$
to $x_j$. Sign-inconsistencies arise when a feedback loop has negative
sign, or the two paths of a feedforward loop from $x_i$ to $x_j$ have
opposite signs. Sign-consistent loops are feedback loops with positive
sign, as well as feedforward loops with the two paths having the same
sign. Similar motifs have been outlined e.g. in \cite{Maa2008}.

To measure how strongly the connectivity is distributed among the three
nodes, the \textit{maximum indegree} is computed for each network:
\begin{equation}
  maxindeg(\A) := \max_{i \in\{1,2,3\}} %
  \left\lbrace \sum_{j \in\{1,2,3\}} {a_{ij}} \right\rbrace ,  \label{eq:maxindeg}
\end{equation}
which is the largest row-sum, thus the $\infty$-norm $|| \A ||_{\infty}$.

\subsection{Robustness measure} \label{sec:meth:rob}

Since the goal of this study is to compare networks that can reproduce
the desired cell types, we computed for each network structure a
robustness measure $\mathcal{R}$ that quantifies if and how well this
structure can generate a set of forward-invariant sets 
as specified in (\ref{eq:specSSs1},\ref{eq:specSSs2}).  We applied the
algorithm introduced in \cite{Breindl2010} which
is summarized next.

In order to explain the underlying idea of this method, let us use the
symbol $\varphi$ to denote both, activation and inhibition
functions. These functions are indexed $\varphi_{i,k}$ such that $i$
denotes the regulated TR, i.e., $x_i$, and $k$ enumerates the
regulators of $x_i$. The number of regulators of $x_i$ is denoted as
$q_i$.

Also, let us measure a perturbation of a monotonous function $\varphi$
as the $l_1$-norm of the difference of the original monotonous
function $\varphi$ and the perturbed monotonous function $\varphi^p$,
i.e.,
\begin{equation}
  \label{eq:perturbation}
  \|\varphi - \varphi^p\|_1=\int_0^\infty |\varphi(x) - \varphi^p(x)|\,\mathrm{d}x.
\end{equation}

Given a set of desired forward-invariant sets $\x^z$, $z=1,\ldots,m$,
the goal of the method is to assign a robustness value $\mathcal{R}$
to the interaction structure itself, i.e., to the matrix $\A$ as in
(\ref{eq:interamat}). This robustness value $\mathcal{R}$ is defined
such that it has the following three properties: (i) If a value
$\mathcal{R}$ is assigned to a system $\A$, there exists a realization of
$\A$ with monotonous functions $\tilde{\varphi}_{i,k}$, such that the
desired sets $\x^z$, $z=1,\ldots,m$, are forward-invariant. (ii) If no
monotonous function is perturbed by more than $\mathcal{R}$, i.e.,
$\forall i,k: \|\tilde{\varphi}_{i,k} - \tilde{\varphi}^p_{i,k}\|_1
\leq \mathcal{R}$, it can be guaranteed that forward-invariance of the
sets $\x^z$, $z=1,\ldots,m$, is maintained. (iii) For every
$\hat{\mathcal{R}} > \mathcal{R}$, there exist perturbed functions
$\hat{\varphi}^p_{i,k}$ such that forward-invariance of the sets $\x^z$,
$z=1,\ldots,m$, is lost, and for at least one index $i,k$ it holds that
$\|\tilde{\varphi}_{i,k}-\hat{\varphi}_{i,k}^p\|_1 =
\hat{\mathcal{R}}$.

The procedure to compute $\mathcal{R}$ for a given network structure
$\A$ and specification $\x^z$, $z=1,\ldots,m$, is outlined
next. To this end, two definitions are given next.
\begin{definition}
  The 3-tuple of pairs of positive real numbers \\ $T_{\nu}=
  \left((x^{\low},\gamma^{\low}), (x^{\high},\gamma^{\high}),(x^{\maxi},\gamma^{\maxi})\right)$
  such that $\gamma^{\low} \leq \gamma^{\high} \leq \gamma^{\maxi}$
  and $x^{\low} \leq x^{\high} \leq x^{\maxi}$ is called tube for
  an activation function.
  Equivalently, the 3-tuple of pairs of positive real numbers \\
  $T_{\mu}= \left((x^{\low},\gamma^{\high}),(x^{\high},\gamma^{\low}),(x^{\maxi},\gamma^{\mini})\right)$
  such that $\gamma^{\mini} \leq \gamma^{\low} \leq \gamma^{\high}$
  and $x^{\low} \leq x^{\high} \leq x^{\maxi}$ is called tube for an
  inhibition function.
\end{definition}

\begin{definition}
  An activation function $\nu$ (inhibition function $\mu$) is said to
  satisfy a tube $T_{\nu}$ ($T_{\mu}$), denoted as $\nu \vDash T_{\nu}$
  ($\mu \vDash T_{\mu}$) if the following inequalities hold.
  \begin{eqnarray}
    \forall x \leq x^{\low}  &: \ \nu(x) \leq \gamma^{\low}\ (\mu(x) \geq \gamma^{\high})%
    \label{eq:tube1}\\
    \forall x \geq x^{\high} &: \ \nu(x) \geq \gamma^{\high}\ (\mu(x) \leq \gamma^{\low})%
    \label{eq:tube2}\\
    \forall x \leq x^{\maxi} &: \ \nu(x) \leq \gamma^{\maxi}\ (\mu(x) \geq \gamma^{\mini})%
    \label{eq:tube3}
  \end{eqnarray}
\end{definition}

\begin{figure}
  \begin{center}
         \includegraphics{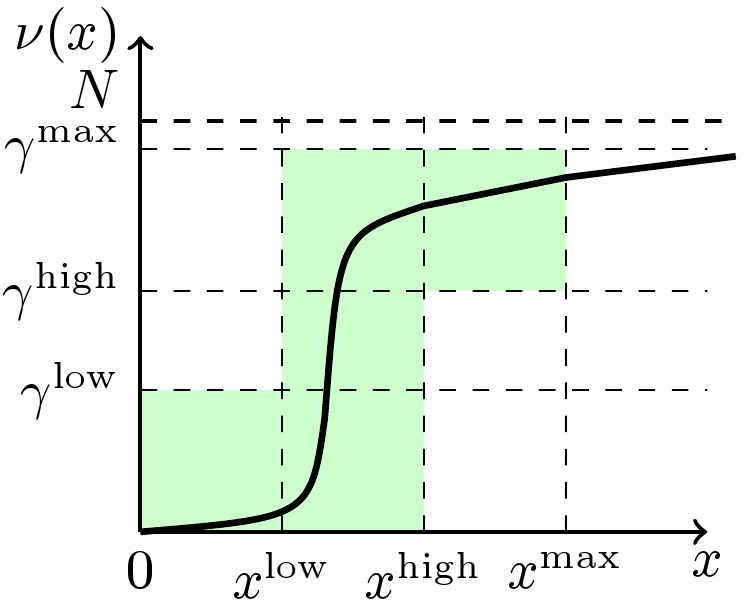}
     \end{center}
  \vspace*{-.4cm}
  \caption{Illustration of a tube for an activation function.}
  \label{fig:tube}
\end{figure}

This means that a tube for a monotonous function restricts the space
where the graph of this function can evolve (see also
Figure~\ref{fig:tube}).  Note that the $x$-values of the tubes are
given by the specification of the forward-invariant sets. It is now
possible to formulate conditions on the $\gamma$-values of the tubes
such that the specified sets are forward-invariant if all functions
lie inside their tubes, i.e.  $\forall i,k: \varphi_{i,k} \vDash
T^{i,k}$, where $T$ can mean $T_{\nu}$ or $T_{\,u}$ as required, and
the tubes are indexed as the according monotonous functions.

\begin{prop}[\cite{Breindl2010}]
  \label{prop:inv_tubes}
  Given a set $\x = \intervx{1}^{l_1} \times \ldots \times
  \intervx{n}^{l_n}$ with $l_i \in \{\low,\high\}$ and given tubes
  $T^{i,k}$ that satisfy the conditions
  \begin{equation}
    \label{eq:tubeconstr3}
    \forall i \in \{1, \ldots n\}:\ %
    -k_i \cdot \underline{x}_{i} + %
    \underline{\gamma}_{i,1} \cdot \ldots \cdot \underline{\gamma}_{i,q_i} \geq 0
  \end{equation}
  where $\underline{x}_i = \min_{x_i \in \intervx{i}^{l_i}} x_i$, and,
  with $x_j$ denoting the argument of $\varphi_{i,k}$,
  \begin{equation*}
    \underline{\gamma}_{i,k} =
    \left\{
      \begin{array}{ll}
        0 & \hspace*{-1.9cm}\mbox{if } %
        \varphi_{i,k} = \nu_{i,k} \wedge 0 \in \intervx{j}^{l_j} \\
        \min \{ \gamma: (x,\gamma) \in T^{i,k} \wedge\; x \in \intervx{j}^{l_j} \}  %
        & \mbox{otherwise}
      \end{array}
    \right.
  \end{equation*}
  and
  \begin{equation}
    \label{eq:tubeconstr4}
    \forall i \in \{1, \ldots n\}:\ %
    -k_i \cdot \overline{x}_{i} + %
    \overline{\gamma}_{i,1} \cdot \ldots \cdot \overline{\gamma}_{i,q_i} \leq 0
  \end{equation}
  where $\overline{x}_i = \max_{x_i \in \intervx{i}^{l_i}} x_i$, and,
  with $x_j$ denoting the argument of $\varphi_{i,k}$,
  \begin{equation*}
    \overline{\gamma}_{i,k} =
    \max \{ \gamma: (x,\gamma) \in T^{i,k} \wedge x_j \in \intervx{j}^{l_j} \}.
  \end{equation*}
  If $\forall i,k:\ \varphi_{i,k} \vDash T^{i,k}$, then the set
  $\x$ is forward-invariant for system~(\ref{eq:system}).
\end{prop}

Starting from this result, the robustness measure can now be
introduced. As forward-invariance for the specified sets can be
guaranteed as long as $\forall i,k: \varphi_{i,k} \vDash T^{i,k}$, the
goal is to find, for each tube $T^{i,k}$, the function
$\tilde{\varphi}_{i,k}$, which is best centered to the tube
$T^{i,k}$. In this context, best centered means that $\tilde{\varphi}_{i,k}$ has
to be perturbed more than any other function $\varphi_{i,k} \vDash
T^{i,k}$ in order to violate at least one of the constraints
(\ref{eq:tube1}-\ref{eq:tube3}) (denoted as $\varphi_{i,k} \nvDash
T^{i,k}$).  The minimal perturbation of this best centered function
$\tilde{\varphi}_{i,k}$ in order to achieve a violation of the tube
constraints is given as the result of the optimization problem
\begin{equation}
  \label{eq:max_robustness_tube}
  \mathcal{R}^{\maxi}(T^{i,k})=\sup_{\varphi_{i,k} \vDash T^{i,k}} %
  \mathcal{R}^{\mini}(\varphi_{i,k},T^{i,k}),
\end{equation}
with
\begin{equation}
  \mathcal{R}^{\min}(\varphi_{i,k},T^{i,k})=\inf_{\varphi^p_{i,k} %
    \nvDash T^{i,k}} \|\varphi_{i,k}-\varphi^p_{i,k}\|_1.
\end{equation}
Then, the robustness measure is obtained by maximizing the smallest value
$\mathcal{R}^\maxi(T^{i,k})$ of all tubes in the network over all tubes $T^{i,k}$ that
satisfy Proposition \ref{prop:inv_tubes}.
\begin{definition}
Given a system (\ref{eq:system}) with unspecified activation and inhibition functions.
The robustness measure $\mathcal{R}$  for the system is defined as
\begin{equation}
  \label{eq:problem1}
  \begin{aligned}
    \mathcal{R}  = & \max_{T^{i,k}} \min_{i,k}  \mathcal{R}^{\maxi}(T^{i,k}) \\
    \mbox{s.t.: } &  \forall \x^z: %
    \mbox{(\ref{eq:tubeconstr3}) and (\ref{eq:tubeconstr4}) hold.}
  \end{aligned}
\end{equation}
\end{definition}
This measure has the interpretation given at the beginning of this
section.  Furthermore, if there exists no realization of $\A$ such
that the desired sets are forward-invariant, the optimization
problem~(\ref{eq:problem1}) is infeasible.  It could furthermore be
shown that for networks where the functions $f_i(x)$ are products of
activation and inhibition functions, problem~(\ref{eq:problem1}) results in a
convex optimization problem \citep{Breindl2010}.


\section{Results} \label{sec:res}

\subsection{General observations}

Constructing all possible 3-node networks with interactions as
specified in (\ref{eq:interamat}), i.e. each interaction having
exactly one value $\in\{-1,0,1\}$, yields $3^{3 \cdot 3} = 19683$
possible networks.  Requiring the networks to be weakly connected
leaves $19008$ networks.  Out of these candidate networks, the
algorithm from Section \ref{sec:meth:rob} identifies $206$ models that
meet the specification (S1), i.e. $1.08\%$ out of all networks; and
$242$ models that are able to reproduce the specification (S2),
i.e. $1.27\%$ out of all networks.

\begin{figure}[!bt]
\begin{center}
\includegraphics[scale=.73]{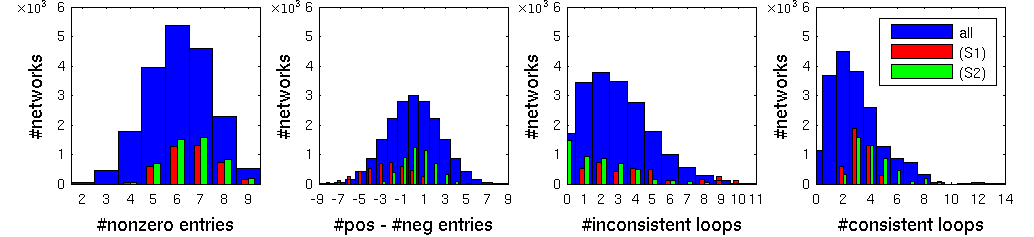}\\
\includegraphics[scale=.73]{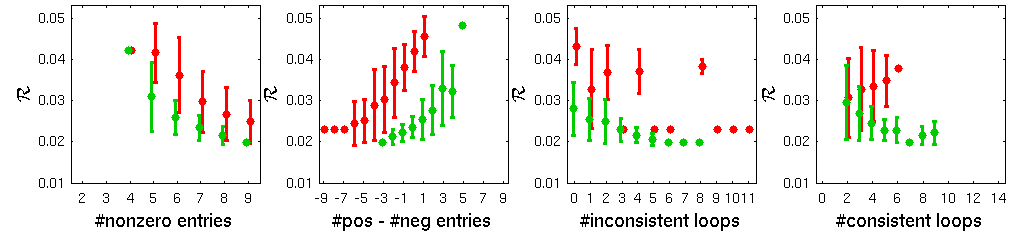}
\end{center}
\vspace*{-.3cm}
\caption{\label{fig:distchars} (Upper row) Distributions of
  characteristic measures of all 19008 possible weakly-connected
  3-node networks (blue), of the 206 models reproducing the
  SSS-specification (S1) (red, $\times 20$), and of the 242 models
  reproducing (S2) (green, $\times 20$).
  (Lower row) Mean robustness values for each characteristic
  measure, over the 206 models reproducing (S1) (red), and over the
  242 models reproducing (S2)
  (green).}
\end{figure}

The distributions of the characteristic measures as outlined in
Section~\ref{sec:meth} are shown in Fig.~\ref{fig:distchars}. 
Models for (S1) (Fig.~\ref{fig:distchars} upper row, red) and models
for (S2) (Fig.~\ref{fig:distchars} upper row, green) show similar
distributions of nonzero entries.  The distribution of the
positive-negative weight for models of (S1) is shifted to the
left, i.e. toward more negative entries, whereas for models of (S2) it
is slightly shifted to the right, i.e. toward more positive entries.
These differences are in accordance with the number of inconsistent
loops which require negative entries: Models for (S1) show a
distribution of inconsistent loops up to high numbers and only
intermediate amount of consistent loops, whereas models for (S2) tend
to low numbers of inconsistencies and more consistent loops.  Also,
models for (S1) (red) are found for all numbers of inconsistent loops,
whereas models for (S2) (green) are restricted to up to 8 inconsistent
loops. In contrast, the (S1)-models are more restricted to certain numbers
of consistent loops than the (S2)-models are.

Summarizing, there are less networks supporting the hypothesis of a
progenitor factor (S1), and they show a tendency
toward negative (inhibiting) interactions and hence
inconsistent loops. The number of networks providing models for the
hypothesis of a differentiation factor (S2) is
higher. On average, these networks have about equal number of
positive and negative entries, thus reducing the amount of
inconsistent loops in favor of consistent loops.

\subsection{Structural properties}

Regarding the network structure, we found that all models that can reproduce
a SSS-specification can be classified leading to sufficient and necessary
conditions on the interaction structure. These findings are summarized in the
following observations. 

\begin{figure}[ht]
\begin{center}
  \includegraphics{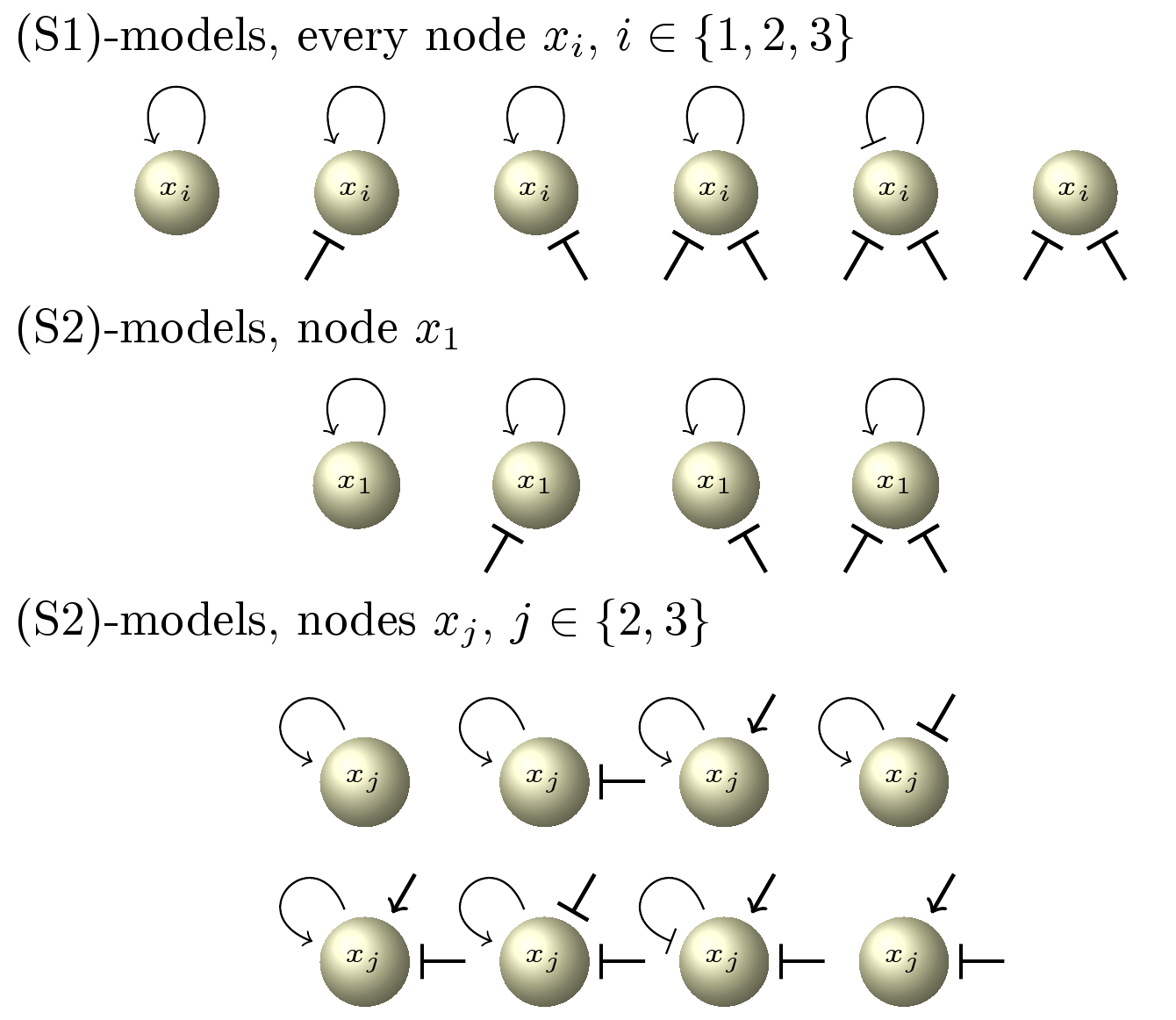}
\end{center}
\caption{Complete enumeration of all ``building blocks''. Networks
  assembled from these building blocks are able to reproduce the
  respective SSS-specification. (Upper row:) Links entering $x_i$ from
  bottom left and right represent links from nodes $x_j$, $j \neq i$
  and $x_k$, $i \neq k \neq j$. (Middle row:) Links entering $x_1$ from
  below left and right represent links from nodes $x_2$ and $x_3$. (Lower row:)
Links entering from top right represent links from $x_1$. Links entering from right
represent links from $x_k$, $k \in \{2,3\}$, $k \neq j$.}
\label{fig:Building_blocks}
\end{figure}

\begin{observation}
  Let $\A$ be an interaction matrix (\ref{eq:interamat}).  If the
  network represented by $\A$ is such that each node $x_i$, $i \in
  \{1,2,3\}$ together with its incoming links is contained in
  Fig.~\ref{fig:Building_blocks}, then there exists a realization of
  $\A$ that can reproduce the SSS-specification (S1). Also the other
  direction holds: For each weakly connected 3-node network that can
  be assembled from nodes $x_{i}$ together with its incoming links
  contained in Fig.~\ref{fig:Building_blocks} there exists a
  realization that can reproduce the SSS-specification (S1).
   \label{obs:propS1}
\end{observation}

An equivalent observation can be made for the network that can reproduce the
SSS-specification (S2).

\begin{observation}
  Let $\A$ be an interaction matrix (\ref{eq:interamat}).  If the
  network represented by $\A$ is such that the node $x_1$ and the
  nodes $x_j$, $j \in \{2,3\}$ together with their incoming links are
  contained in Fig.~\ref{fig:Building_blocks}, then there exists a
  realization of $\A$ that can reproduce the SSS-specification
  (S1). Also the other direction holds:  For each weakly connected
  3-node network that can be assembled from nodes $x_{1}$ and $x_j$,
  $j \in \{2,3\}$, together with their incoming links contained in
  Fig.~\ref{fig:Building_blocks} there exists a realization that can
  reproduce the SSS-specification (S2).
   \label{obs:propS2}
\end{observation}

From Fig.~\ref{fig:Building_blocks} it can also be explained why the
positive-negative weight of models for SSS-specification (S1) tends
toward a larger number of negative entries: All interactions between
two different nodes in the network need to be negative
(inhibiting). Only self-loops are allowed to be activating.  For
SSS-specification (S2) on the other side also activating links from
$x_{1}$ to the other nodes are possible.

Using these ``building blocks'' as listed in
Fig.~\ref{fig:Building_blocks} it is possible to construct all
possible 3-node models that can, for an appropriate choice of
activation and inhibition functions, reproduce the SSS-specification
(S1) or (S2).  To see possible applications of our approach, note that
a switching module for mesenchymal stem cell differentiation based on
literature data and analyzed in detail in \cite{Schittler2010}
satisfies specification (S1) and can indeed be composed from the
building blocks presented in Fig.~\ref{fig:Building_blocks}.  By
applying the algorithm from Section~\ref{sec:meth:rob}, it is in
principle possible to compute the building blocks for any
SSS-specification. The algorithm is also directly applicable to larger
networks.

\subsection{Robustness properties}

The mean robustness value $\R$ for all characteristic measures is
shown in Fig.~\ref{fig:distchars} (lower row). Two important
observations can be made: There is a clear negative correlation of
$\mathcal{R}$ with increasing nonzero entries, and a clear positive
correlation of $\mathcal{R}$ with increasing positive-negative
weight. Both observations are true independent of the specific
SSS-specifications (red, green).  In contrast, no clear correlation of
$\R$ versus the number of inconsistent or consistent loops can be
detected.

\begin{figure}[!b]
\begin{center}
\includegraphics{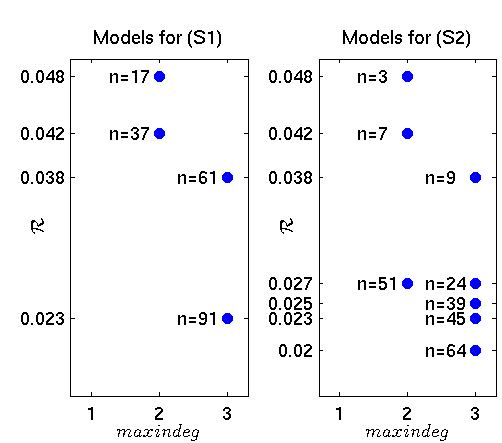}
\end{center}
\caption{\label{fig:RV_maxindeg} The robustness value $\R$ versus the
  $maxindeg$ for all models. For each $(maxindeg,\R)$, there are $n$
  models which have these values of $maxindeg, \R$.  (Left) For
  specification (S1), there are four different robustness values, each
  of them uniquely assigned to a $maxindeg$ value.  (Right) For
  specification (S2), there are seven different robustness values. The
  assignment of a given $\R$ to a $maxindeg$ value is not unique.  }
\end{figure}

It is stated by \cite{Kwon2008} that perturbations on nodes involved
in fewer (more) than the average number of feedback loops have a lower
(higher, respectively) impact on the overall network robustness.
Thus, we investigated whether there is a similar relationship between
the $maxindeg$ and the robustness value $\R$.  The results of our
analysis are shown in Fig.~\ref{fig:RV_maxindeg}.
Among all (S1)-reproducing models, only four different robustness
values have been assigned while for the (S2)-reproducing models seven
different robustness values have been computed.  Furthermore, the (S1)-models
are more abundant in higher robustness values, whereas the majority of
(S2)-models has a lower robustness value
(cf. Fig.~\ref{fig:RV_maxindeg}).  That is, models for hypothesis (S2)
require a finer ``tuning'' than models for (S1), biologically pointing
toward more vulnerable regulatory networks.


Yet, despite these differences, the similarities between the two
distributions are eminent which suggests a strong relation between the
network's structural properties and their robustness: Independent of
the specification (S1) or (S2), networks with a higher $maxindeg$ tend
to have lower $\mathcal{R}$ values. This means that, the more incoming
interactions a node has, the more vulnerable the network gets to
perturbations of these interaction.  This is in accordance with the
findings of \cite{Kwon2008} about network structure and their notion
of robustness.

\section{Conclusion} \label{sec:con}

The main focus of this work was to analyze which networks consisting
of three transcriptional regulators are in principle able to reproduce
a set of required stable steady states, to search for interaction
patterns that are necessary in order to meet these requirements, and
to compute and compare the robustness properties of these networks.
Still, the investigations raise new questions and possible future work,
which we point out briefly in this concluding outlook.


Sign-inconsistencies make a system non-monotone, and the more
inconsistent loops are contained in a network, the ``farther'' the
system is from monotonicity.  \cite{Maa2008} argue that intracellular
systems will be ``close-to-monotone''. \cite{Kwon2008} report that a
higher number of positive feedback loops and a smaller number of
negative feedback loops result in a higher robustness.
These arguments are in accordance with the result here that positive
(negative) entries increase (decrease) the robustness of a network.
But in contrast to the number of positive / negative entries, the
number of inconsistent and consistent loops does not show a clear
effect on the robustness as defined here. What is more, inconsistent
loops are even required for some SSS-specifications to be met.
These studies use different notions of robustness, and their
interrelation should be worth further investigations.

Furthermore, one might argue that especially in cell differentiation,
stimulus inputs and transitions between stable steady states rather
than just the existence of stable steady states play an important
role. However, although these aspects were not considered in this
study, the approach in this paper provides a first selection step of
networks with respect to one necessary condition (out of several),
namely by checking whether the required stable steady states can be
reproduced.
These aspects provide plenty of opportunities for separate
investigations, and we hope to address these issues in future work.

\section*{Achknowledgement}
\begin{small}  The authors would like to thank for financial support from the
  German Research Foundation (DFG) within the Cluster of Excellence in
  Simulation Technology (EXC 310/1) at the University of Stuttgart,
  and from the German Federal Ministry of Education and Research
  (BMBF) within the SysTec program (grant nr. 0315506A), and from the
  Helmholtz Society within the CoReNe project.  D.S. acknowledges
  financial support by The MathWorks Foundation of Science and
  Engineering.\end{small}


\begin{thebibliography}{xx}  

\begin{small}
  
\bibitem[Breindl et~al.(2010)]{Breindl2010} C. Breindl, S. Waldherr
  and F. Allg{\"o}wer.  \newblock A robustness measure for the
  stationary behavior of qualitative gene regulation networks.
  \newblock Proceedings of the 11th symposium on computer applications in
  biotechnology (CAB), pp. 36--41, 2010.
  
\bibitem[Chaves et~al.(2008)]{Chaves2008} M. Chaves, T. Eissing and
  F. Allg{\"o}wer.  \newblock Bistable Biological Systems: A
  Characterization Through Local Compact Input-to-State
  Stability. \newblock IEEE Transactions on Automatic Control, Special
  Issue on Systems Biology, vol.~53, pp. 87--100, 2008.

\bibitem[Foster et~al.(2009)]{Foster2009}
  D.V. Foster, J.G. Foster, S. Huang and S.A. Kauffman.
  \newblock A model of sequential branching in hierarchical cell fate determination.
  \newblock Journal of Theoretical Biology, vol.~260, pp. 589--597, 2009.
  
\bibitem[Huang et~al.(2007)]{Huang2007}
  S. Huang, Y.-P. Guo, G. May and T. Enver.
  \newblock Bifurcation dynamics in lineage-commitment in bipotent progenitor cells.
  \newblock Developmental Biology, vol.~305, pp. 695--713, 2007.
 
\bibitem[Kwon et~al.(2008)]{Kwon2008}
  Y.-K. Kwon and K.-H. Cho.
  \newblock Quantitative analysis of robustness and fragility in biological networks based on feedback dynamics.
  \newblock Bioinformatics, vol.~24, pp. 987--994, 2008.

\bibitem[Ma'ayan et~al.(2008)]{Maa2008}
  A. Ma'ayan, A. Lipshtat, R. Iyengar and E.D. Sontag.
  \newblock Proximity of intracellular regulatory networks to monotone systems.
  \newblock IET Systems Biology, vol.~2, pp. 103--112, 2008.
  
\bibitem[MacArthur et~al.(2008)]{MacA2008}
  B.D. MacArthur, C.P. Please and R.O.C. Oreffo.
  \newblock Stochasticity and the molecular mechanisms of induced pluripotency.
  \newblock PLoS ONE, vol.~3, e3086, 2008.
  
\bibitem[MacArthur et~al.(2009)]{MacA2009}
  B.D. MacArthur, A. Ma'ayan and I.R. Lemischka.
  \newblock Systems biology of stem cell fate and cellular reprogramming.
  \newblock Nature Reviews Molecular Cell Biology, vol.~10, pp. 672--681, 2009.

\bibitem[Narula et~al.(2010)]{Narula2010} J. Narula, A.M. Smith,
  B. Gottgens and O.A. Igoshin.  \newblock Modeling reveals
  bistability and low-pass filtering in the network module determining
  blood stem cell fate.  \newblock PLoS Computational Biology, vol.~6,
  e1000771, 2010.
  
\bibitem[Peltier and Schaffer(2010)]{Peltier2010}
  J. Peltier and D.V. Schaffer.
  \newblock Systems biology approaches to understand stem cell fate choice.
  \newblock IET Systems Biology, vol.~4, pp. 1--11, 2010.
  
\bibitem[Roeder and Glauche(2006)]{Roeder2006} I. Roeder and
  I. Glauche.  \newblock Towards an understanding of lineage
  specification in hematopoietic stem cells: A mathematical mode for
  the interaction of transcription factors GATA-1 and PU.1.  \newblock
  Journal of Theoretical Biology, vol.~241, pp. 852--865, 2006.

\bibitem[Schittler et~al.(2010)]{Schittler2010} D.Schittler,
  J. Hasenauer, F. Allg\"ower, and S.Waldherr.  \newblock Cell
  differentiation modeled via a coupled two-switch regulatory network.
  \newblock Chaos, vol.~20(4):045121, 2010.
  
\bibitem[Xiong and Ferrell(2003)]{Xiong2003} W. Xiong and
  J.E. Ferrell, Jr..  \newblock A positive-feedback-based bistable
  memory module that governs a cell fate decision.  \newblock Nature,
  vol.~426, pp. 460--465, 2003.


\end{small} 
\end{thebibliography}

\end{document}